# Microscopic annealing process and its impact on superconductivity in *T′*-structure electron-doped copper oxides

Hye Jung Kang[1,2,3], Pengcheng Dai[1,4], Branton J. Campbell[5], Peter J. Chupas[6], Stephan Rosenkranz[6], Peter L. Lee[7], Qingzhen Huang[2], Shiliang Li[1], Seiki Komiya[8], and Yoichi Ando[8]

[1]*Department of Physics and Astronomy, The University of Tennessee, Knoxville, Tennessee 37996-1200, USA*

[2]*NIST Center for Neutron Research, National Institute of Standards and Technology, Gaithersburg, MD 20899-8562, USA*

[3]*Department of Materials Science and Engineering, University of Maryland, College Park, MD 20742-2115, USA*

[4]*Neutron Scattering Sciences Division, Oak Ridge National Laboratory, Oak Ridge, Tennessee 37831-6393, USA*

[5]*Department of Physics and Astronomy, Brigham Young University, Provo, Utah 84602, USA*

[6]*Materials Science Division, Argonne National Laboratory, Argonne, Illinois 60439, USA*

[7]*Advanced Photon Source, Argonne National Laboratory, Argonne, Illinois 60439, USA*

[8]*Central Research Institute of Electric Power Industry, Komae, Tokyo 201-8511, Japan*

**High-transition-temperature superconductivity arises in copper oxides when holes or electrons are doped into the $CuO_2$ planes of their insulating parent compounds. While hole-doping quickly induces metallic behavior and superconductivity in many cuprates, electron-doping alone is insufficient in materials such as $R_2CuO_4$ ($R$ is Nd, Pr, La, Ce, etc.), where it is necessary to anneal an as-grown sample in a low-oxygen environment to remove a tiny amount of oxygen in order to induce superconductivity. Here we show that the microscopic process of oxygen reduction repairs Cu deficiencies in the as-grown materials and creates oxygen vacancies in the stoichiometric $CuO_2$ planes, effectively reducing disorder and providing itinerant carriers for superconductivity. The resolution of this long-standing materials issue suggests that the fundamental mechanism for superconductivity is the same for electron- and hole-doped copper oxides.**

The parent compounds of the high-transition-temperature (high-$T_c$) copper-oxide superconductors are antiferromagnetic (AF) Mott insulators composed of two-dimensional $CuO_2$ planes separated by charge reservoir layers[1-3]. When holes are doped into these planes, the static long-range AF order is quickly destroyed and the lamellar copper-oxide materials become metallic and superconducting over a wide hole-doping range. In the case of electron-doped materials such as the $T'$-structured $R_2CuO_4$ ($R$ is Nd, Pr, La, Ce, etc.), electron-doping alone is insufficient, and annealing the as-grown sample in a low oxygen environment to remove a tiny amount of oxygen is necessary to induce superconductivity[2,3]. Previous work[4-23] suggests that oxygen reduction may influence mobile carrier concentrations[7], decrease disorder/impurity scattering[8,10,11,23], or suppress the long-range AF order[16,17,22]. However, the microscopic process of oxygen reduction, its effect on the large electron-hole phase diagram asymmetry and mechanism of superconductivity[2,3] are still unknown. Here we use x-ray and neutron scattering data, combined with chemical and thermo-gravimetric analysis measurements in the electron-doped $Pr_{0.88}LaCe_{0.12}CuO_4$ to show that the microscopic process of oxygen reduction is to repair Cu deficiencies in the as-grown materials[12,13] and to create oxygen vacancies in the stoichiometric $CuO_2$ (refs. 16,17,22), effectively repairing disorder in the $CuO_2$ planes and providing itinerant carriers for superconductivity.

The role of the reduction process in the superconductivity of electron-doped high-$T_c$ copper oxides has been a long-standing unsolved problem. For the hole-doped cuprates, low doping levels (e.g. 5%) entirely suppress AF order and superconductivity appears over a wide range of hole concentrations (from 6% to 30%). In the case of $T'$ structured electron-doped superconductors, doping alone by substituting the trivalent ions $R^{3+}$ in $R_2CuO_4$ with tetravalent $Ce^{4+}$ is insufficient to induce superconductivity and the as-grown materials such as $Nd_{2-x}Ce_xCuO_{4\pm\delta}$ (NCCO), $Pr_{2-x}Ce_xCuO_{4\pm\delta}$ (PCCO), and $Pr_{0.88}LaCe_{0.12}CuO_4$ (PLCCO) are semiconducting with static long-range AF order[4-22]. In these materials, superconductivity (maximum $T_c$ ~25 K) is achieved only when the samples within a relatively narrow dopant range ($0.1 \leq x \leq 0.18$) are annealed in a reducing atmosphere [2-23]. The reduction process suppresses the AF order most severely at higher Ce-doping levels above $x \geq 0.1$ (ref. 14). This so-called "materials issue" in $T'$ structured electron-doped superconductors raises the question whether the fundamental mechanism for superconductivity is the same for electron- and hole-doped superconductors. For example, while the symmetry of the superconducting order parameter in hole-doped superconductors is generally believed to be $d$-wave[24], there has been no consensus on the nature of the superconducting order parameter in electron-doped materials[25]. A resolution of the "materials issue" surrounding the annealing process should allow us to focus on the intrinsic similarities and differences between the electron and hole-doped superconductors.

Originally, Tokura et al.[2,3] suggested that an as-grown sample had stoichiometric oxygen. The reduction process removes oxygen in the $CuO_2$ plane (making oxygen content less than 4), and therefore introduces mobile electrons into the system[2]. This microscopic oxygen-reduction mechanism was based on the resistivity values from undoped $Nd_2CuO_4$ samples, which were found to decrease dramatically upon quenching from 1150 °C to room temperature. While experiments using a thermo-gravimetric apparatus (TGA) confirm that the reduction process removes some of the oxygen[4-6], this simple model is inconsistent with observations that as-grown samples have excess oxygen per Cu atom[7-9] and that superconductivity is not achieved at any Ce (*i.e.*, electron) doping levels without subsequent oxygen reduction.

Using neutron diffraction, resistivity and Hall-effect measurements, Greene and co-workers argue that the annealing process increases the number of mobile charge carriers[7] or decreases impurity scattering by removing apical oxygens (proposed oxygen just above and below the Cu atoms of the $CuO_2$ planes) which should be absent in the ideal *T'* structure of electron-doped materials[8,10,11,23]. In this latter scenario, the presence of a small amount of randomly doped apical oxygen in the as-grown materials induces localization of doped electrons and thus prohibits superconductivity. The role of the oxygen reduction process would then be to remove the excess apical oxygen and increase the mobility of doped electrons to allow metallic behavior and superconductivity[8,10,11,23].

Although the reduction process is widely believed to involve removal of the apical oxygen, such notion has recently been challenged by Raman, infrared transmission, and ultrasound studies of electron-doped materials[16,17,22]. Instead of removing apical oxygen, these experiments on NCCO and PCCO suggest that the reduction process removes the oxygen in the $CuO_2$ plane [O(1), see Fig. 1a] at high Ce-doping ($x \geq 0.1$) and the apical oxygen at low Ce-doping[16,17]. In such a circumstance, the in-plane oxygen defect created by the reduction would be responsible for superconductivity by destroying the long-range antiferromagnetism and increasing the mobility of charge carriers[16,17,22]. Because the antiferromagnetism is much less affected at low Ce doping levels[14], superconductivity is not achieved for $x < 0.1$ even though the samples are reduced.

While these current microscopic scenarios appear plausible, they all miss a key ingredient of the problem, that is, none of these models can explain why the appearance of superconductivity after oxygen reduction is also intimately related to the creation of an epitaxially grown impurity phase of cubic $R_2O_3$ (refs. 15, 18-21). Experimentally, it was found that about 1%~2% of the impurity phase, absent in the as-grown materials, appears with superconducting phase after the oxygen reduction process[18-21]. Although not explicitly demonstrated, the $R_2O_3$ impurity phase may disappear again when annealed samples are oxygenated to eliminate superconductivity[15,21]. Therefore, superconductivity in electron-doped materials might be intimately related to the appearance of the impurity phase.

Here we employed systematic x-ray and neutron diffraction measurements, combined with chemical and TGA measurements in single crystals and powders of PLCCO to elucidate the microscopic origin of the annealing process. Because of Cu evaporation during the high-temperature synthesis process, a small amount of Cu deficiency (~1%) was found in the stoichiometrically prepared as-grown NCCO[12,13]. According to Kang *et al.*[20], as-grown samples of PLCCO may also be Cu deficient. To test this possibility, single crystals of PLCCO (space group *I4/mmm*, $a$ = 3.98 Å, $c$ = 12.3 Å) were grown using the traveling-solvent floating-zone technique. These included four samples prepared from the same as-grown batch: as-grown nonsuperconducting (ag-NSC), reduced superconducting (r-SC), oxygenated non-superconducting (o-NSC), and re-reduced superconducting (r2-SC) crystals.

As shown in previous work[18-21], annealing necessary to produce superconductivity also induces epitaxial intergrowths of the cubic $R_2O_3$ impurity phase. The $R_2O_3$ has a cubic structure with space group *Ia3* and lattice parameter $a_c = 2\sqrt{2}a \sim c/1.1$, where the "$c$" subscript indicates "cubic". The Miller indices of the *T'* and impurity phases are related as $(H, K, L) = [(H_c - K_c)/4, (H_c + K_c)/4, L = 1.1L_c]$. Since the lattice parameter of $R_2O_3$ matches the in-plane lattice parameter of the *T'* phase but is ~10% less than the *c*-axis lattice parameter of the *T'* phase, Bragg reflections from the impurity phase should occur at commensurate positions in the $(H, K)$ reciprocal space plane but at $L$=1.1, 2.2, and etc. For odd values of $L$, structure factors of the impurity phase become non-zero at (1/2, 0) and (1/4, 1/4) type positions[18-21]. Figure 2 summarizes x-ray diffraction mesh scans of the $(H, K)$ reciprocal space plane of PLCCO at $L$= 1.1 taken from the ag-NSC, r-SC, and r2-SC single crystal samples. The data from the ag-NSC (Fig. 2a) reveals thermal diffuse scattering associated fundamental Bragg peaks in the $T'$ phase. In the SC sample (Figs. 2b), a new set of peaks develop at (1/2, 0) and (1/4, 1/4) type positions, confirming the presence of the impurity phase.

To test the reversibility of the impurity phase, we carried out x-ray powder diffraction measurements on portions of the ag-NSC, r-SC, and o-NSC single crystal samples of PLCCO that were crushed into fine powders. In addition, r-SC and o-NSC ceramic samples of PLCCO were prepared and analyzed for comparison with the crushed single-crystal samples. Figure 3 shows a portion of the x-ray powder patterns from each of these samples. The two SC samples clearly reveal the $(2, 2, 2)_c$ reflection of the $R_2O_3$ impurity phase at $2\theta = 8.7°$. This peak is conspicuously absent in the three NSC samples. Furthermore, the impurity phase appears again in re-reduced r2-SC single crystal (Fig. 2c). These results conclusively show that the impurity phase grows epitaxially parallel to the $a$-$b$ plane of the $T'$ cuprate structure, and it is reversibly created upon oxygen reduction to the SC state and destroyed upon oxidation to the NSC state. Thus, the oxygen reduction process is completely reversible and intimately related to the appearance of the impurity phase and superconductivity in PLCCO.

To understand this phenomenon, consider an annealing process that causes a slightly Cu-deficient $T'$ material to phase-separate into a Cu-perfect $T'$ majority phase and a Cu-free minority phase. As shown in Figs. 1a-d, removing the Cu atoms from a single $CuO_2$ plane results in a locally primitive pseudo-cubic oxygen sublattice with 50% of the cubic interstices filled by rare-earth atoms. When viewed from the [1,1,0] direction, this structure cannot be distinguished from that of fluorite, though the rare-earth sublattice is not yet face centered. The creation of one oxygen vacancy for each Cu atom removed, followed by a vacancy-assisted one-interstice displacement of the nearby rare-earth atoms on one side of the affected sheet then produces the familiar face-centered rare-earth sublattice of fluorite -- a 25% anion-deficient fluorite in fact. This 25% anion vacancy concentration marks the extreme limit of stability for the fluorite structure and is only known to occur in the vacancy-ordered structure type commonly referred to as the "*C*-type sesquioxide" or the "bixbyite" structure. If we assume that ag-NSC samples are composed of $R_2Cu_{1-f}O_{4-\alpha}$, where $f$ and $\alpha$ are Cu and oxygen deficiencies respectively, the annealing process can be written as $R_2Cu_{1-f}O_{4-\alpha} = (1-f)(R_2CuO_{4-\beta}) + f(R_2O_3) + (-\alpha + \beta + f(1-\beta))(O)$, where $\beta$ is the oxygen deficiency level in the superconducting part of $R_2CuO_4$.

The appeal of such a structural reorganization lies in the fact that the observed $R_2O_3$ impurity phase has precisely the *C*-type sesquioxide structure. The Cu atoms removed from the affected sheet, rather than irreversibly leaving the sample, are free to fill vacancies in other layers, thus "repairing" most of the Cu vacancies in the majority *T'* phase. And just as oxygen removal appears to drive the phase separation, subsequent high-temperature oxidation is assumed to drive a redistribution of Cu vacancies throughout the sample and the concomitant elimination of the epitaxial sheets of $R_2O_3$. Such an annealing process is entirely reversible, and benefits from the fact that the *T'* structure is stable against Cu deficiencies due to the three-dimensionally interconnected rare earth-oxygen network[5] that includes rather than being interrupted by $CuO_2$ planes.

To establish any Cu deficiencies in the samples, we performed inductively-coupled plasma (ICP) atomic-emission spectroscopy analysis on ag-NSC, an r-SC, and an o-NSC PLCCO crystals. Similar measurements were also carried out on ceramic PLCCO samples. The ICP results are summarized in table **I** and indicate that all samples have a small Cu deficiency. It is worth noting here that the ICP analyses provide no information on oxygen content, and that they average over the *T'* cuprate and Cu-free impurity components of the mixed-phase SC samples. They should, therefore, be taken as a qualitative rather than a quantitative indicator of Cu deficiency.

To quantitatively determine the volume fraction of the impurity phase, we carried out the Rietveld analysis of the x-ray powder patterns using the GSAS software as summarized in Table **II**. The volume fraction of the impurity phase in the SC samples is about 1.6%. The composition of PLCCO was also refined, yielding average Cu deficiencies of 3.5% in the NSC samples and 1.4% in the SC samples. The difference in Cu deficiency between the NSC and SC samples is then quite close to the refined volume fraction of the impurity phase (~1.6%), further confirming the relationship between Cu vacancies and impurity phase formation.

The use of an x-ray refinement to determine the absolute occupancy of each unique atom in PLCCO is difficult because of its coupling to the atomic displacement parameters (*i.e.* thermal factors). To obtain better occupancies, neutron powder-diffraction data were collected using two different wavelengths, after which combined double-wavelength Rietveld refinements were performed. The key results are summarized in Table **III**, the most interesting of which are the differences in Cu and oxygen occupancies between the SC and the NSC samples. The occupancy of in-plane oxygen O(1) (Fig. 3 inset) is also significantly more deficient in the SC samples (~1%). The Cu occupancy is 1.2% to 2.3% deficient in the NSC samples, but recovers full occupancy in the SC samples. Once again, the difference in Cu-deficiency levels between SC and NSC samples is comparable to the volume fraction of the impurity phase obtained from x-ray scattering. Because the impurity phase does not contain Cu, the missing Cu in an NSC sample should be consistent with the volume fraction of the impurity phase in an SC sample. Furthermore, because the impurity phase forms via oxygen reduction as illustrated in Fig. 1, one should be able to predict the amounts of the oxygen loss using the results of Table **III** and compare them to actual TGA oxygen-loss measurements for each sample. This is indeed the case as shown in the Methods section.

Using the refinement results in Table **III**, one can also normalize the Cu occupancies to one in order to obtain Cu:O ratios. The SC samples have normalized oxygen occupancies less than 4 in the chemical formula unit (e.g. $CuO_{4-\beta}$ with $\beta \approx$ 0.06), while the NSC samples have normalized oxygen occupancies greater than 4 (e.g. $CuO_{4-\alpha}$ with $-\alpha \approx$ 0.02 to 0.07). Based on these oxygen occupancies and bond lengths[26], we estimate the valence of Cu to be 1.66 for the SC samples and 1.72 in the NSC samples. The decrease of Cu valence in the SC samples indicates that the SC samples have slightly more doped electrons than the NSC samples. This is mainly due to a smaller number of oxygen atoms per Cu atom, since there is little difference in the bond length between the SC and the NSC samples. Although the annealing process only marginally affects the effective electron doping level, it is large enough to increase electron mobility and decrease the resistivity in the annealed undoped samples[2].

The Cu deficiency in as-grown electron-doped materials provides a natural explanation of its electronic properties, effectively weakening the electron-doping effect of Ce by increasing numbers of oxygens and therefore adding holes per Cu atom. In the present model, the annealing process diminishes any Cu deficiency and therefore promotes the electron-doping effect of Ce, which also suppresses the AF Néel temperature ($T_N$). Our recent systematic neutron diffraction measurements on PLCCO as a function of annealing process showed that the $T_N$ of PLCCO drops from 186 K in an ag-NSC sample to <600 mK in an r-SC sample ($T_c$ = 24 K)[27,28].

From recent Hall coefficient measurements, Gauthier *et al.*[29] argue that the annealing process does not affect the carrier concentration but rather affects the mobility of the carriers. Since random Cu and oxygen vacancies in the $CuO_2$ plane break the translational symmetry of the $CuO_2$ plane, they can act as impurity centers to localize the doped electrons. The observed increase in charge carrier mobility would then appear to be the result of eliminating the Cu vacancies and repairing the $CuO_2$ planes of the $T'$ structure. Of course, the creation of oxygen defects in the $CuO_2$ plane as a byproduct of the oxygen reduction process (Table **III**) may also be necessary to suppress the AF order and increase the electron mobility[16,17,22], therefore facilitating superconductivity in the electron-doped cuprates.

Differences in the phase diagrams[2,3] and the electronic properties[24,25] of the electron and hole-doped cuprate superconductors have been obscured by the "materials issue" surrounding the annealing process in the electron-doped samples. The microscopic oxygen-reduction process presented here removes much of the mystery and will need to be taken into account in order to resolve any intrinsic differences between electron and hole-doped superconductivity. Angle resolved photoemission (ARPES) measurements on NCCO (ref. 30) and PLCCO (ref. 31) have recently determined the Fermi surface topology and the superconducting gap function in electron-doped materials. However, the quality of the sample surfaces studied has not been conclusively established, and one must now consider the possible effect of an exposed NSC $R_2O_3$ surface in future ARPES and scanning tunneling microscopy experiments. Furthermore, it should also be possible to grow higher-quality single-crystal $T'$ structured copper oxides by adding excess Cu in the starting materials to eliminate Cu deficiencies and subsequent $R_2O_3$ contamination.

Because of the availability of single crystals of the $T'$ structured copper oxides, the vast majority of transport, structural, magnetic, and electronic-property measurements of electron-doped superconductors have been obtained from these materials[9-22,25,27-31]. For other electron-doped high-$T_c$ superconductors with different crystal structure types, such as the infinite-layer $Sr_{1-x}La_xCuO_2$ family[32-34], the microscopic process of inducing superconductivity is still unclear[33,34]. There, the large single crystals that have made numerous single-crystal X-ray and neutron scattering experiments possible in the $T'$ materials[18-21] are not currently available. In case of the $T'$ structured materials, however, we can return our focus to the electronic properties of the $CuO_2$ planes themselves in the search for a mechanism of superconductivity.

**Methods**

Single crystals of PLCCO were grown using the traveling-solvent floating-zone technique. Four samples (ag-NSC, r-SC, o-NSC, and r2-SC) were then prepared from the same as-grown batch of crystals[27]. Single-crystal x-ray diffraction experiments were performed at room temperature with 30 keV x-rays at the BESSRC 11-ID beamline of the Advanced Photon Source (APS) at Argonne National Laboratory. Synchrotron x-ray powder diffraction measurements were performed at a wavelength of 0.49582 Å on the 32-ID beamline at APS. Reciprocal lattice vectors have been defined relative to the $T'$ cuprate lattice of PLCCO and are labeled as $(H, K, L) = (Q_x\,a, Q_y\,a, Q_z$

$c)/2\pi$, where $\mathbf{Q} = (Q_x, Q_y, Q_z)$ is the scattering wave vector in units of Å$^{-1}$. Rietveld refinements were performed on powder diffraction data to obtain site occupancies and phase fractions. Because the rare earth atoms (Pr, La, Ce) share a single site and very similar x-ray scattering amplitudes, their relative ratios were fixed according to the nominal synthetic composition, and the total rare-earth occupancy was normalized to 2. Because of the weak relative x-ray scattering amplitude of oxygen, the oxygen occupancies were also fixed at 100%. Thus, only the Cu occupancy and the impurity phase fraction were refined (Table **II**).

Neutron diffraction measurements at 10 K were performed at the BT-1 beamline of the NIST Center for Neutron Research, using two different monochromators: Cu(311) with $\lambda$=1.5403 Å and Ge(733) with $\lambda$=1.1968 Å. The use of two wavelengths helped to decouple site occupancies from thermal factors. The Cu(311) monochromator has higher intensity and achieves good resolution at mid-range scattering angles, whereas the Ge(733) has less than 1/3 of the intensity of Cu(311), but achieves better resolution at high scattering angles. Combined double-wavelength Rietveld refinements took full advantage of these data (Table **III**). All samples were refined in the *I4/mmm* space group using the following atomic positions: *R*(rare earth): 4e[0 0 z], Cu: 2a[0 0 0], O(1): 4c[0 1/2 0], O(2): 4d [0 1/2 1/4]. The occupancies were normalized to yield a Pr+Ce occupancy of one. On the rare earth site, only the occupancy of La was refined, since Pr and Ce have very similar neutron scattering lengths. The occupancy of O(2) was also fixed to its ideal value. The Cu bond valences were calculated[26] using the site occupancies and bond-lengths of each refined structure. The impurity peaks are too weak to observe in neutron powder diffraction patterns because the rare-earth elements have much smaller neutron-scattering amplitudes relative to Cu and O than was the case for x-rays.

TGA oxygen-loss measurements of an ag-NSC crystal were performed while annealing the samples between 24 °C and 830 °C at a vacuum pressure of 5×10$^{-5}$ mbar. Assuming $R_2Cu_{1-f}O_{4-\alpha}$ and $R_2CuO_{4-\beta}$ (R=Pr, La, and Ce) before and after the reduction step respectively, we find $-\alpha + \beta + f(1-\beta) = 0.041$ using Table **III**, consistent with TGA measurement of the oxygen loss of 0.04. The consideration of small rare-earth deficiencies does not influence the conclusions.

**Acknowledgements**

We thank Klaus Attenkofer for the help with the synchrotron X-ray scattering experiments at the 11-ID beamline. The x-ray and neutron scattering work is supported in part by the US National Science Foundation with Grant No. DMR-0453804 and by an award from Research Corporation. The PLCCO single crystal growth at UT is supported by the US DOE BES under contract No. DE-FG02-05ER46202. ORNL is supported by the US DOE Grant No. DE-AC05-00OR22725 through UT/Battelle LLC. Work at Argonne was supported by the US DOE under contract No. DE-AC02-06CH11357. The part of the work done in Japan was supported by the Grant-in-Aid for Science provided by the Japan Society for the Promotion of Science.

Correspondence and Request for Materials should be addressed to H.J.K (hkang@nist.gov) or P.D. (daip@ornl.gov).

**Figure captions:**

**Figure 1 Crystal structures, magnetic susceptibilities, and TGA measurements. a,** Typical $T'$ structure ($I4/mmm$) of PLCCO consisting of $CuO_2$ sheets separated by thin sheets of rare-earth oxide (Pr,La,Ce)O. **b,** $T'$ structure of as-grown PLCCO with random Cu vacancies in the $CuO_2$ planes. **c,** A modified $T'$ structure in which Cu atoms have been removed from one layer, leaving behind a thin slab of the anion-deficient fluorite impurity phase: $R_2O_3$ = (Pr,La,Ce)$_2O_3$. **d,** The vacancy-ordered $C$-type sesquioxide structure of the $R_2O_3$ impurity phase (space group $Ia3$). **e,** Magnetic susceptibilities of the r-SC single crystal and ceramic powder samples as a function of temperature. **f,** Oxygen losses during the reduction process measured with a TGA. The mass change between 500 °C and room temperature is mainly due to the intrinsic properties of the TGA instrument in vacuum.

**Figure 2 Single-crystal x-ray diffraction mesh scans in the ($H$, $K$) plane at $L$ = 1.1.** PLCCO samples exhibit thermal diffuse scattering around intense Bragg reflections. **a,** The ag-NSC sample shows no evidence of the $R_2O_3$ impurity phase. **b,** The r-SC sample, prepared by reducing the ag-NSC sample, exhibits peaks at (1/2, 0) and (1/4, 1/4) positions (marked by solid squares) associated with epitaxial intergrowths of the $R_2O_3$ impurity phase. **c,** The r2-SC sample shows identical behavior as r-SC.

**Figure 3 X-ray powder diffraction patterns from crushed single-crystal and ceramic powder samples.** The SC samples reveal the (2, 2, 2)$_c$ Bragg reflection of the $R_2O_3$ impurity phase at 2θ=8.7 °. The inset illustrates the $T'$ crystal structure and atom labels used in Rietveld refinements.

**Tables:**

**Table I Nominal PLCCO metal compositions via ICP analysis.** ICP standard solutions were prepared by dissolving $Pr_6O_{11}$, $La_2O_3$, $CeO_2$, and $CuO$ powders in $HNO_3$. Each rare-earth powder was baked to remove moisture before measuring its mass. For each PLCCO sample, 30 mg of material was dissolved into 100 mL of $HNO_3$ (1mol/L) acid. The compositions are normalized to give a Pr composition of 0.88.

| Samples | Nominal composition from ICP measurement |
|---|---|
| ag-NSC | $Pr_{0.88}La_{0.95}Ce_{0.10}Cu_{0.90}$ |
| r-SC | $Pr_{0.88}La_{0.95}Ce_{0.10}Cu_{0.90}$ |
| o-NSC | $Pr_{0.88}La_{0.95}Ce_{0.10}Cu_{0.90}$ |
| r-SC ceramic | $Pr_{0.88}La_{1.01}Ce_{0.10}Cu_{0.95}$ |

**Table II Nominal PLCCO compositions and phase fractions from x-ray powder diffraction.** The 1.6% volume fraction of the impurity phase in the SC samples is much more sensitive to x-rays than neutrons because of rare earth atoms' relatively-high x-ray scattering amplitude. This value is comparable to the difference in Cu deficiency levels between the NSC and SC samples.

| Samples | Nominal composition from x-ray scattering | Volume fraction of $(Pr,La,Ce)_2O_3$ |
|---|---|---|
| ag-NSC | $Pr_{0.88}LaCe_{0.12}Cu_{0.962(3)}O_4$ | |
| o-NSC ceramic | $Pr_{0.88}LaCe_{0.12}Cu_{0.962(3)}O_4$ | |
| o-NSC | $Pr_{0.88}LaCe_{0.12}Cu_{0.970(4)}O_4$ | |
| r-SC | $Pr_{0.88}LaCe_{0.12}Cu_{0.983(4)}O_4$ | 0.0163 (1.63 %) |
| r-SC ceramic | $Pr_{0.88}LaCe_{0.12}Cu_{0.990(4)}O_4$ | 0.0157 (1.57 %) |

**Table III Neutron refinement results of PLCCO samples.** All samples were refined in space group *I4/mmm*. Occupancies are shown in terms of atoms per chemical formula unit. The statistical deviations of refined parameters are included in parentheses. The principle difference between the SC and the NSC samples is a significantly lower Cu deficiency and a lower bond valence in the SC samples.

| Parameter | r-SC | ag-NSC | r-SC ceramic | o-NSC ceramic |
|---|---|---|---|---|
| $n(Pr)$ | 0.88 | 0.88 | 0.88 | 0.88 |
| $n(La)$ | 0.962(8) | 0.978(8) | 0.958(8) | 0.960(8) |
| $n(Ce)$ | 0.12 | 0.12 | 0.12 | 0.12 |
| $z$ | 0.35312(5) | 0.35273(5) | 0.35302(5) | 0.35277(4) |
| $u_{11}=u_{22}$ (Å$^2$) | 0.0056(2) | 0.0051(2) | 0.0053(2) | 0.0049(2) |
| $u_{33}$ (Å$^2$) | 0.0084(4) | 0.0078(4) | 0.0072(3) | 0.0070(3) |
| | | | | |
| $n(Cu)$ | 1.001(6) | 0.977(6) | 1.001(6) | 0.988(5) |
| $u_{11}=u_{22}$ (Å$^2$) | 0.0056(3) | 0.0051(3) | 0.0064(3) | 0.0063(2) |
| $u_{33}$ (Å$^2$) | 0.0129(5) | 0.0106(5) | 0.0109(4) | 0.0100(4) |
| | | | | |
| $n(O(1))$ | 1.95(1) | 1.98(1) | 1.94(1) | 1.97(1) |
| $u_{11}$ (Å$^2$) | 0.0091(4) | 0.0090(4) | 0.0105(4) | 0.0100(3) |
| $u_{22}$ (Å$^2$) | 0.0062(4) | 0.0119(4) | 0.0073(4) | 0.0091(3) |
| $u_{33}$ (Å$^2$) | 0.0155(4) | 0.0126(4) | 0.0140(4) | 0.0129(3) |
| | | | | |
| $n(O(2))$ | 2.00 | 2.00 | 2.00 | 2.00 |
| $u_{11}=u_{22}$ (Å$^2$) | 0.0071(2) | 0.0068(2) | 0.0085(2) | 0.0074(2) |
| $u_{33}$ (Å$^2$) | 0.0146(4) | 0.0121(3) | 0.0126(3) | 0.0120(3) |
| **Ratio(Cu:O)** | CuO$_{3.95}$ | CuO$_{4.07}$ | CuO$_{3.94}$ | CuO$_{4.02}$ |
| | | | | |
| RE-O(1) (Å) | 2.6922(4) | 2.6948(4) | 2.6933(4) | 2.6952(4) |
| RE-O(2) (Å) | 2.3640(3) | 2.3609(3) | 2.3636(3) | 2.3615(3) |
| O(1)-O(1) (Å) | 2.81995(3) | 2.81921(3) | 2.82001(2) | 2.81923(2) |
| O(1)-O(2) (Å) | 3.07875(4) | 3.07805(3) | 3.07940(3) | 3.07980(3) |
| Cu-O(1) (Å) | 1.99401(2) | 1.99348(2) | 1.99405(1) | 1.99350(2) |
| $R(F^2)$ | 5.6% | 6.2% | 4.5% | 4.1% |
| $wR$ | 5.4% | 5.4% | 5.9% | 5.62% |
| $\chi^2$ | 1.377 | 1.383 | 1.223 | 1.248 |
| **Cu bond valence (e.u.)** | **1.66** | **1.73** | **1.66** | **1.71** |

**Figures:**

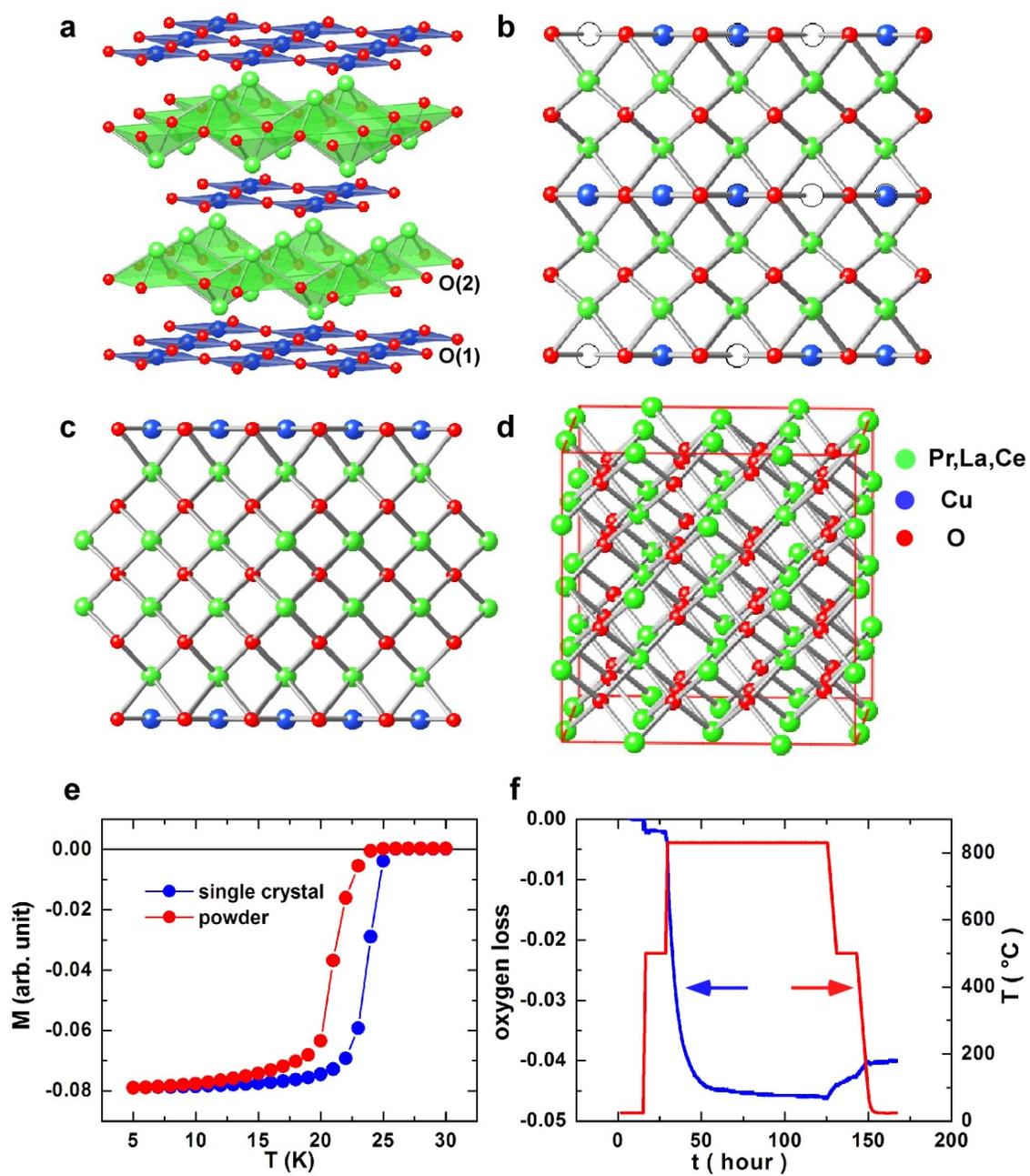

Fig. 1

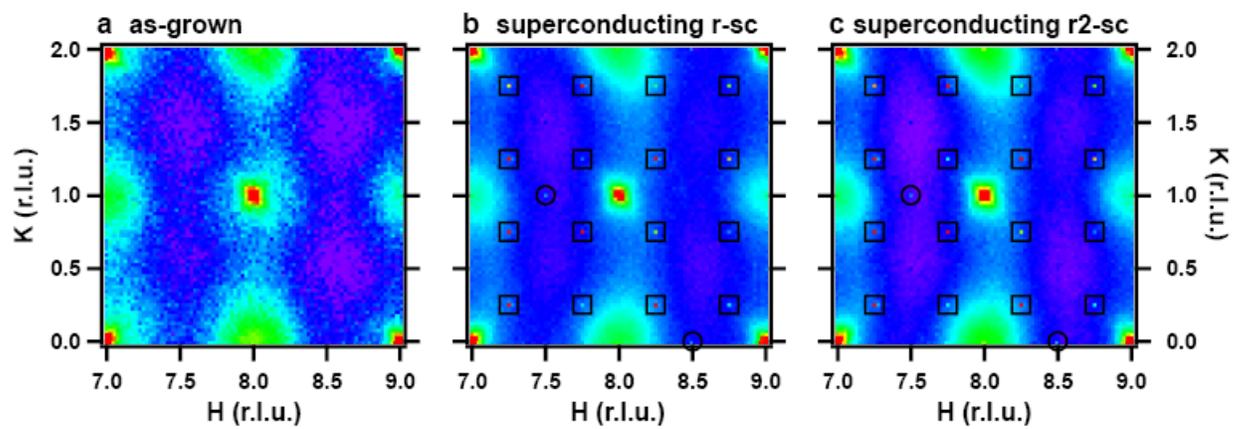

**Figure 2**

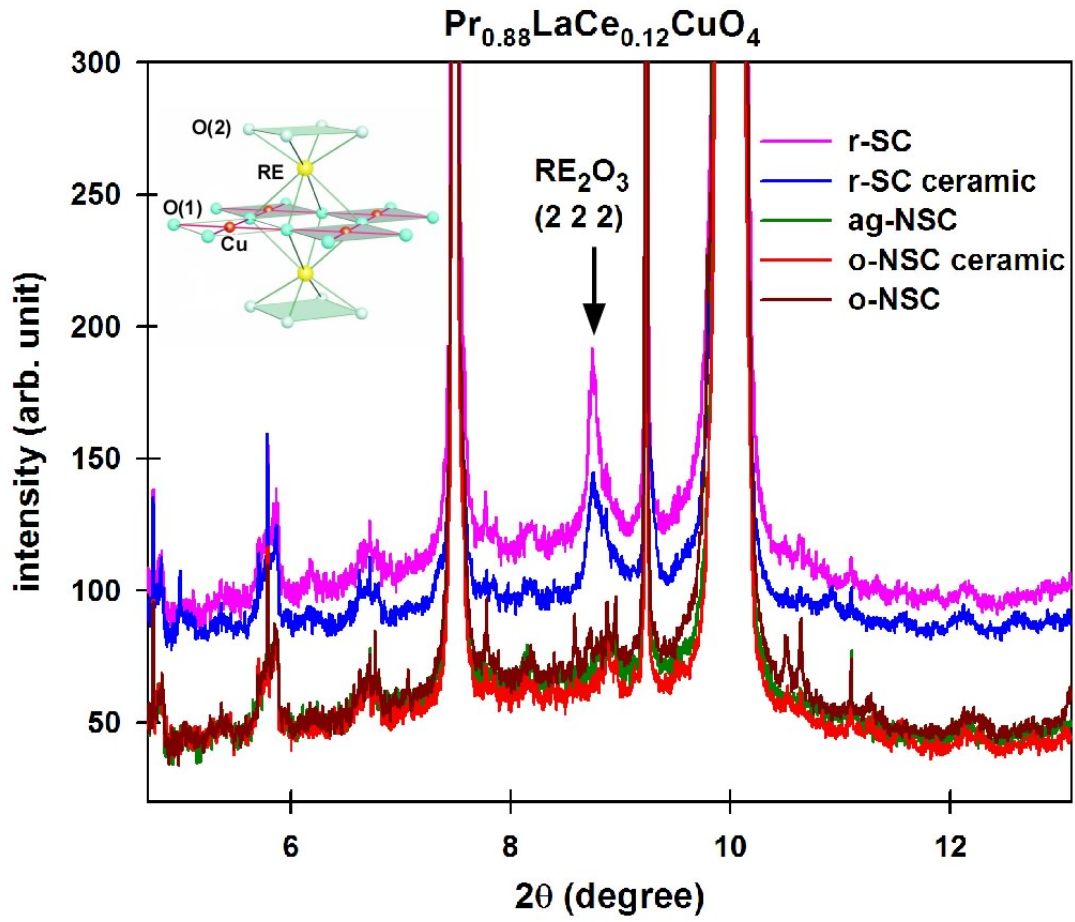

Fig. 3